\documentclass[a4paper,11pt]{article}

\usepackage{fullpage}

\usepackage{booktabs} % For formal tables

\usepackage[ruled]{algorithm}
\usepackage{algpseudocode}

\usepackage{latexsym}
\usepackage{url}
\usepackage{xspace}
\usepackage{color}
\usepackage{graphicx}

\usepackage{xspace}

\newcommand{\mpicommsplit}{\textsf{MPI\_\-Comm\_\-split}\xspace}
\newcommand{\mpicommcreate}{\textsf{MPI\_\-Comm\_\-create}\xspace}
\newcommand{\mpicommcreategroup}{\textsf{MPI\_\-Comm\_\-create\_\-group}\xspace}
\newcommand{\mpibcast}{\textsf{MPI\_\-Bcast}\xspace}
\newcommand{\mpiallreduce}{\textsf{MPI\_\-Allreduce}\xspace}
\newcommand{\mpialltoall}{\textsf{MPI\_\-Alltoall}\xspace}
\newcommand{\mpialltoallv}{\textsf{MPI\_\-Alltoallv}\xspace}
\newcommand{\MPIMAX}{\texttt{MPI\_\-MAX}\xspace}

\newcommand{\openmpilib}{\texttt{OpenMPI 3.0.0}\xspace}

\newcommand{\Cqsort}{\texttt{qsort()}\xspace}

\newcommand{\eg}{e.g.\@\xspace}

\newcommand{\leaveout}[1]{#1}

\usepackage{tikz}
\usetikzlibrary{arrows.meta}
\tikzset{>={Classical TikZ Rightarrow[width=1.7mm,length=1mm]}}

\title{Parallel Quicksort without Pairwise Element Exchange}
\author{Jesper Larsson Tr\"aff\\
TU Wien, Faculty of Informatics, Research Group Parallel Computing\\
Favoritenstrasse 16/191-4\\
1040 Vienna, Austri\\
\texttt{traff@par.tuwien.ac.at}}
\date{February 6th, 2018, revision 15th October 2018}

\begin{document}
\maketitle

\begin{abstract}
  Quicksort is an instructive classroom approach to parallel sorting
  on distributed memory parallel computers with many opportunities for
  illustrating specific implementation alternatives and tradeoffs with
  common communication interfaces like MPI. The (two) standard
  distributed memory Quicksort implementations exchange partitioned
  data elements at each level of the Quicksort recursion. In this
  note, we show that this is not necessary: It suffices to distribute
  only the chosen pivots, and postpone element redistribution to the
  bottom of the recursion. This reduces the total volume of data
  exchanged from $O(n\log p)$ to $O(n)$, $n$ being the total number of
  elements to be sorted and $p$ a power-of-two number of processors,
  by trading off against a total of $O(p)$ additional pivot element
  distributions. Based on this observation, we describe new,
  \emph{exchange-free} implementation variants of parallel Quicksort
  and of Wagar's HyperQuicksort.

  We have implemented the discussed four different Quicksort
  variations in MPI, and show that with good pivot selection,
  Quicksort without pairwise element exchange can be significantly
  faster than standard implementations on moderately large
  problems. For smaller input sizes, standard and exchange-free
  variants can be combined to exploit the exchange-free variant as
  subproblems become large enough relative to the number of
  processors. On a medium sized high-performance cluster,
  exchange-free parallel Quicksort achieves an absolute speed-up of
  about 650 on $1024$ processes when sorting $10^8$ floating point
  numbers. In a weak-scaling experiment on $8192$ processes, the
  exchange-free variants are consistently about a factor 1.3 faster
  than their standard counterparts when the number of elements per
  process exceeds $500\,000$.  The combination of standard and
  exchange-based variant never performs worse than either, and for a
  medium number of elements per process, better than both.
\end{abstract}

%\begin{CCSXML}
%\end{CCSXML}

\maketitle

\section{Introduction}

Quicksort~\cite{Hoare62} is often used in the classroom as an example
of a sorting algorithm with obvious potential for parallelization on
different types of parallel computers, and with enough obstacles to
make the discussion instructive. Still, distributed memory parallel
Quicksort is practically relevant (fastest) for certain ranges of
(smaller) problem sizes and numbers of
processors~\cite{AxtmannSanders17,AxtmannWiebigkeSanders18}.

This note presents two new parallel variants of the Quicksort scheme:
select a pivot, partition elements around pivot, recurse on two
disjoint sets of elements. A distributed memory implementation needs
to efficiently parallelize both the pivot selection and the
partitioning step in order to let the two recursive invocations
proceed concurrently. In standard implementations, partitioning
usually involves exchanging elements between neighboring processors in
a hypercube communication pattern. We observe that this explicit
element exchange is not necessary; it suffices instead to distribute
the chosen pivots over the involved processors. This leads to two new
\emph{exchange-free} parallel Quicksort variants with a cost tradeoff
between element exchange and pivot distribution. We discuss
implementations of the two variants using the \emph{Message-Passing
  Interface} (MPI)~\cite{MPI-3.1}, and compare these to standard
implementations of parallel Quicksort.  Experiments on a medium scale
cluster show that this can be faster than the standard pairwise
exchange variants when the number of elements per process is not too
small. The two approaches can be combined for a smooth transition
between exchange-based and exchange-free Quicksort.

Using MPI terminology, we let $p=2^k$ denote the number of MPI
\emph{processes} that will be mapped to physical processor(core)s. For
the Quicksort variants discussed here, $p$ must be a power of two. MPI
processes are \emph{ranked} consecutively, $i=0,\ldots,p-1$.  We let
$n$ denote the total number of input elements, and assume that these
are initially distributed evenly, possibly
randomized~\cite{AxtmannSanders17} over the $p$ processes, such that
each process has roughly $n/p$ elements. Elements may be large and
complex and hence expensive to exchange between processes; but all
have a key from some ordered domain with a comparison function $<$ that
can be evaluated in $O(1)$ time. For each process, input and output
elements are stored consecutively in process local arrays. For each
process, the number of output elements should be close to the number
of input elements, but actual load balance depends on the quality of
the selected pivots. The output elements for each process
must be sorted, and all output elements for process $i$ must be smaller than or
equal to all output elements of process $i+1$, $0\leq i<p-1$.

\section{Standard, distributed memory Quicksort and HyperQuicksort}

Standard, textbook implementations of parallel Quicksort for
distributed memory systems work roughly as
follows~\cite{Quinn03,GramaKarypisKumarGupta03,LanMohamed92,SundarMalhotraBiros13,Wagar87}.
%Input of $n$ elements are distributed roughly evenly over the $p$ processes
%such that each has about $n/p$ elements.
%The number of processes is
%assumed to be a power of two, and communication follows a hypercube
%pattern.
A global pivot is chosen by some means and distributed over
the processes, after which the processes all perform a local
partitioning of their input elements. The processes pair up such that
process $i$ is paired with process $i\oplus p/2$ (with $\oplus$
denoting bitwise exclusive or), and process pairs exchange data
elements such that all elements smaller than (or equal to) the pivot
end up at the lower ranked process, and elements larger than (or
equal) to the pivot at the higher ranked process. After this, the set
of processes is split into two groups, those with rank lower than
$p/2$ and those with larger rank. The algorithm is invoked recursively
on these two sets of processes, and terminates with a local sorting
step when each process belongs to a singleton set of processes.

Assuming that pivots close to the median element can be effectively
determined, the communication volume for the element exchange over all
recursive calls is $O(n\log p)$. With linear time communication costs,
the exchange time per process is $O(n/p \log p)$.  Global pivot
selection and distribution is normally done by the processes locally
selecting a (sample of) pivot(s) and agreeing on a global pivot by
means of a suitable collective operation. If we assume the cost for
this to be $O(\log p+s)$ where $s, s\geq 1$, is the size of the pivot
sample per process, the total cost for the pivot selection becomes
$O(\log p(\log p+s))$. Some textbook implementations simply use the
local pivot from some designated process which is distributed by an
\mpibcast operation~\cite{Quinn03,GramaKarypisKumarGupta03}; others
use local pivots from all processes from which a global pivot closer
to the median is determined by an \mpiallreduce-like
operation~\cite{AxtmannSanders17,SiebertWolf11}. Either of these take
time $O(\log p+s)$ in a linear communication cost model, see,
\eg~\cite{Traff09:twotree}. Before the recursive calls, the set of MPI
processes is split in two which can conveniently be done using the
collective \mpicommcreate operation, and the recursive calls simply
consist in each process recursing on the subset of processes to which
it belongs. Ideally, \mpicommcreate\footnote{It can be assumed that
  both \mpicommcreate and \mpicommcreategroup are faster than the
  alternative \mpicommsplit in any reasonable MPI library
  implementation; if not, a better implementation of \mpicommcreate
  can trivially be given in terms of \mpicommsplit. For evidence, see,
  \eg~\cite{AxtmannWiebigkeSanders18}.} takes time $O(\log p)$. At the
end of the recursion, each process locally sorts (close to) $n/p$
elements. The best, overall running time for this parallel Quicksort
implementation becomes $O(n/p \log p+n/p \log(n/p)+\log^2 p)=O(n/p
\log n+\log^2 p)$ assuming a small (constant) sample size $s$ is used,
with linear speed-up over sequential $O(n \log n)$ Quicksort when
$n/p$ is in $\Omega(\log n)$. We refer to this implementation variant
as standard, \emph{parallel Quicksort}.

Wagar~\cite{Wagar87} observed that much better pivot selection would
result by using the real medians from the processes to determine the
global pivots. In this variation of the Quicksort scheme, the
processes \emph{first sort} their $n/p$ elements locally, and during
the recursion keep their local elements in order. The exact, local
medians for the processes are the middle elements in the local arrays,
among which a global pivot is selected and distributed by a suitable
collective operation. As above, this can be done in $O(\log p)$ time.
The local arrays are split into two halves of elements smaller (or
equal) and larger (or equal) than the global pivot. Instead of having
to scan through the array as in the parallel Quicksort implementation,
this can be done in $O(\log(n/p))$ time by binary search. Processes
pairwise exchange small and large elements, and to maintain order in
the local arrays, each process has to merge its own elements with
those received from its partner in the exchange. The processes then
recurse as explained above. The overall running time of $O(n \log
n+\log^2 p)$ is the same as for parallel Quicksort. Wagar's name for
this Quicksort variant is \emph{HyperQuicksort}.
Wagar~\cite{Wagar87}, Quinn~\cite{Quinn89,Quinn03}, Axtmann and
Sanders~\cite{AxtmannSanders17} and others show that HyperQuicksort
can perform better than parallel Quicksort due to the possibly better
pivot selection. A potential drawback of HyperQuicksort is that the
process-local merge step can only be done after the elements from the
partner process have been received. In contrast, in parallel
Quicksort, the local copying of the elements that are kept at the
process can potentially be done concurrently (overlapped) with the
reception of elements from the partner process.
%or can be avoided entirely by modifying the local partition step to
%work from several array segments. 

\leaveout{
For completeness, the two standard parallel Quicksort implementation
variants are shown as Algorithm~\ref{alg:standardqsort} and
Algorithm~\ref{alg:sortfirstqsort} in the appendix.
}

\section{Exchange-free, parallel Quicksort}

We observe that the partitioning step can be done without actually
exchanging any input elements. Instead, it suffices to distribute the
pivots, and postpone the element redistribution to the end of the
recursion.

\begin{algorithm}
\caption{Exchange-free, per process Quicksort of elements
  in $n$-element array $a$ for $p=2^k$.}
\label{alg:exchangefreeqsort}
\begin{algorithmic}[1]
\Procedure{ExchangeFreeQsort}{$a,n$}
\State $k\gets p$
\State $as[0]\gets a$ \Comment First segment is whole array $a$
\State $an[0]\gets n$ \Comment of $n$ elements
\Repeat
\State $j\gets 0$ \Comment Segment count
\For{$i=0,k,2k,\ldots,p-k$}
\State \Comment Local pivot selection for segment $i$
\State $x[j]\gets\mbox{\Call{local-Choose}{$as[i],an[i]$}}$
\State $j\gets j+1$
\EndFor
\State \Comment Global consensus on all $j$ pivots
\State $x'[0,\ldots,j-1]\gets\mbox{\Call{global-Choose}{$x[0,\ldots,j-1]$}}$ 
\State $j\gets 0$
\For{$i=0,k,2k,\ldots,p-k$}
\State $n_0\gets\mbox{\Call{Partition}{$as[i],an[i],x'[j]$}}$
\State \Comment{$as[i][0,n_0-1]\leq x'[j], as[i][n_0,an[i]-1]\geq x'[j]$}
\State $as[i+k/2]\gets as[i]+n_0$
\State $an[i+k/2]\gets an[i]-n_0$
\State $an[i]\gets n_0$
\State $j\gets j+1$
\EndFor
\State $k\gets k/2$
\Until{$k=1$}
\State \Call{Alltoall}{$as[0,\ldots,p-1],an[0,\ldots,p-1],bs[0,\ldots,p-1],bn[0,\ldots,p-1]$}
\State $m\gets\sum_{i=0}^{p-1}bn[i]$
\State $b\gets bs[0]$ \Comment Consecutive $bs[i]$ segments
\State \Call{local-Qsort}{$b,m$} \Comment Load imbalance $|m-n|$
\EndProcedure
\end{algorithmic}
\end{algorithm}

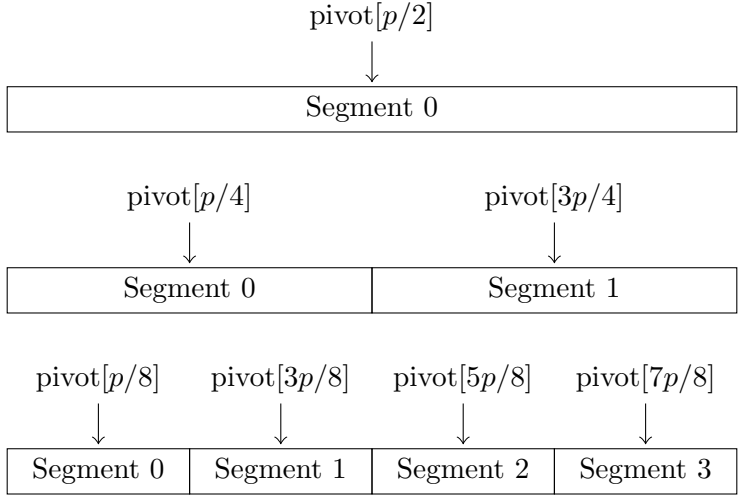
\begin{figure}
\begin{center}
\begin{tikzpicture}[scale=0.60]
%\draw (0,8) rectangle (16,9) node[pos=.5] {Segment 0};
\draw (0,8) rectangle node {Segment 0} (16,9);
\draw[->] (8,10) node () [above] {$\mathrm{pivot}[p/2]$} -- (8,9.1);
%\draw node (16,10) {Iteration 0};

\draw (0,4) rectangle node {Segment 0} (8,5);
\draw (8,4) rectangle node {Segment 1} (16,5);
%\draw (8,5.1) -- (8,3.9);
\draw[->] (4,6) node () [above] {$\mathrm{pivot}[p/4]$} -- (4,5.1);
\draw[->] (12,6) node () [above] {$\mathrm{pivot}[3p/4]$} -- (12,5.1);

\draw (0,0) rectangle node {Segment 0} (4,1);
\draw (4,0) rectangle node {Segment 1} (8,1);
\draw (8,0) rectangle node {Segment 2} (12,1);
\draw (12,0) rectangle node {Segment 3} (16,1);
mpicommsp%\draw (8,1.1) -- (8,-0.1);
%\draw (4,1.1) -- (4,-0.1);
%\draw (12,1.1) -- (12,-0.1);
\draw[->] (2,2) node () [above] {$\mathrm{pivot}[p/8]$} -- (2,1.1);
\draw[->] (6,2) node () [above] {$\mathrm{pivot}[3p/8]$} -- (6,1.1);
\draw[->] (10,2) node () [above] {$\mathrm{pivot}[5p/8]$} -- (10,1.1);
\draw[->] (14,2) node () [above] {$\mathrm{pivot}[7p/8]$} -- (14,1.1);
\end{tikzpicture}
\end{center}
\caption{The first three partitioning iterations, from the viewpoint
  of a single MPI process, assuming for the sake of
  illustration perfect pivot selection. Before the start of iteration
  $i, i\geq 0$, the local array of approximately $n/p$ elements is
  divided into $2^i$ segments. In iteration $i$, all $p$ processes
  agree on $2^i$ new pivots. Each stores the pivots in
  $\mathrm{pivot}[i p/2^{i+1}]$ for $i=0,\ldots,2^i$, and partitions
  the $2^i$ segments for the next iteration.} 
\label{fig:partitioning}
\end{figure}

Algorithm~\ref{alg:exchangefreeqsort} shows how this is realized for
the standard parallel Quicksort algorithm. The idea is best described
iteratively. Before iteration $i, i=0,\ldots,\log_2 p-1$, each
process maintains a partition of its elements into $2^i$ segments
with all elements in segment $j$ being smaller than (or equal to) all
elements in segment $j+1, j=0,\ldots,2^i-2$. In iteration $i$, pivots
for all $2^i$ segments are chosen locally by the processes, and by a
collective communication operation they agree on a global pivot for
each segment.
%This can be done by an all-reduction operation
%with each processor contributing $2^i$ local pivots.
The processes then locally partition their segments, resulting in
$2^{i+1}$ segments for the next iteration. The process is illustrated
in Figure~\ref{fig:partitioning}. After the $\log_2 p$ iterations,
each process has $p$ segments with the mentioned ordering property,
which with good pivot selection each contain about $n/p^2$
elements. By an all-to-all communication operation, all $j$th segments
are sent to process $j$, after which the processes locally sort their
received, approximately $n/p$ elements. Note that no potentially
expensive process set splitting is necessary as was the case for the
standard Quicksort variations. Also note that the algorithm as
performed by each MPI process is actually oblivious to the process
rank. All communication is done by process-symmetric, collective
operations.

If pivot selection is done by a collective \mpibcast or \mpiallreduce
operation, the cost over all iterations will be $O(\log^2 p+p)$, since
in iteration $i$, $2^i$ pivots need to be found and each collective
operation takes $O(\log p+2^i)$ time~\cite{Traff09:twotree}. A small
difficulty here is that some processes' segments could be empty (if
pivot selection is bad) and for such segments no local pivot candidate
is contributed. This can be handled using an all-reduction operation that
reduces only elements from non-empty processes, or by relying on the
reduction operator having a neutral element for the processes not
contributing a local pivot.

This Quicksort algorithm variant can be viewed as a sample sort
implementation (see discussion in~\cite{JaJa00} and,
\eg~\cite{HarschKaleSolomonik18}) with sample key selection done by
the partitioning iterations. No communication of elements is necessary
during the sampling process, only the pivots are distributed in each
iteration. In each iteration, all processes participate which can make
it possible to select better pivots that the standard Quicksort
variations, where the number of participating processes is halved in
each recursive call.

Compared to standard, parallel Quicksort, the $\log_2 p$ pivot
selection and exchange terms of $O(n/p + \log p)$ are traded for a
single $O(n/p + p)$ term accounting for the all-to-all distribution at
the end of the partitioning, and an $O(p+\log^2 p)$ term for the pivot
selection over all $\log_2 p$ iterations, thus effectively saving a
logarithmic factor on the expensive element exchange. The
total running time becomes $O(n/p \log n+\log^2 p+p)$ which means that
this version scales worse than the standard Quicksort
variants with linear speed-up when $n/p$ is in $\Omega(p/\log n)$.

\section{Exchange-free HyperQuicksort}

\begin{algorithm}
\caption{Exchange-free, per process HyperQuicksort
  of elements in $n$-element array $a$ for $p=2^k$.}
\label{alg:sortfirstexchangefreeqsort}
\begin{algorithmic}[1]
\Procedure{ExchangeFreeHyperQsort}{$a,n,i\in\{0,\ldots,p-1\}$}
\State \Call{local-Qsort}{$a,n$}
\State $k\gets p$
\State $as[0]\gets a$ \Comment First segment is whole array $a$
\State $an[0]\gets n$ \Comment of $n$ elements
\Repeat
\State $j\gets 0$ \Comment Segment count
\For{$i=0,k,2k,\ldots,p-k$}
\State \Comment Local pivot selection for segment $j, x[j]=as[an[j]/2]$
\State $x[j]\gets\mbox{\Call{local-Choose}{$as[i],an[i]$}}$
\State $j\gets j+1$
\EndFor
\State \Comment Global consensus on all $2j$ pivots
\State $x'[0,\ldots,j-1]\gets\mbox{\Call{global-Choose}{$x[0,\ldots,j-1]$}}$ 
\State $j\gets 0$
\For{$i=0,k,2k,\ldots,p-k$}
\State $n_0\gets\mbox{\Call{Split}{$as[i],an[i],x'[j]$}}$
\State \Comment{$as[i][0,n_0-1]\leq x'[j], as[i][n_0,an[i]-1]\geq x'[j]$}
\State $as[i+k/2]\gets as[i]+n_0$
\State $an[i+k/2]\gets an[i]-n_0$
\State $an[i]\gets n_0$
\State $j\gets j+1$
\EndFor
\State $k\gets k/2$
\Until{$k=1$}
\State \Call{Alltoall}{$as[0,\ldots,p-1],an[0,\ldots,p-1],bs[0,\ldots,p-1],bn[0,\ldots,p-1]$}
\State $m\gets\sum_{i=0}^{p-1}bn[i]$
\State $b\gets bs[0]$ \Comment Consecutive $bs[i]$ segments
\State \Call{multiway-Merge}{$bn[0,\ldots,p-1],bs[0,\ldots,p-1],c$}
\EndProcedure
\end{algorithmic}
\end{algorithm}

The observation that actual exchange of partitioned elements is not
necessary also applies to the HyperQuicksort
algorithm~\cite{AxtmannSanders17,Quinn03,Wagar87}.  The exchange-free
variant is shown as Algorithm~\ref{alg:sortfirstexchangefreeqsort}. In
iteration $i,i=0,\log_2 p-1$, each process chooses $2^i$ (optimal)
local pivots from each of the $2^i$ segments; each of these local
pivots are just the middle element of the sorted segment. With a
collective operation, $2^i$ global pivots are selected, and each
process performs a split of its segments by binary search for the
pivot. Iterations thus become fast, namely $O(\log^2 p+p+p\log (n/p))$
time for the collective all-reduce for global pivot selection of
$O(\log^2p +p)$ time over all iterations, and $O(p\log (n/p))$ for the
$2^i$ binary searches per iteration. At the end, an (irregular)
all-to-all operation is again necessary to send all $j$th segments to
process $j$. Each of the $p$ segments received per process are
ordered, therefore a multiway merge over $p$ segments is necessary to
get the received elements into sorted order. This takes $O(\log
p\ n/p)$ time.

Since no element exchanges are done in this algorithm and
Algorithm~\ref{alg:exchangefreeqsort} before the all-to-all exchange,
the amount of work for the different processes remains balanced over
the iterations (assuming that all processes have roughly $n/p$
elements to start with).
%However, in contrast to Algorithm~\ref{alg:standardqsort} and
%Algorithm~\ref{alg:sortfirstqsort}, the global collective operations
%introduce a loosely synchronized behavior over the $\log_2 p$
%iterations.

\section{Concrete Implementation and Experimental Results}

We have implemented all four discussed parallel Quicksort variants
with MPI~\cite{MPI-3.1} as discussed. For the process local sorting,
we use the standard, C library \Cqsort function. Speed-up is also
evaluated relative to \Cqsort.

In the experimental evaluation we seek to compare the standard
parallel Quicksort variants against the proposed exchange-free
variants. For that purpose we use inputs where (almost) optimal pivots
can be determined easily. Concretely, each local pivot is selected by
interpolation between the maximum and the minimum element found in a
sample of size $s$. For the HyperQuicksort variants, where input
elements are kept in order, process local minimum and maximum elements
are simply the first and last element in the input element array. For
the standard Quicksort variants, process local maximum and minimum
elements are chosen from a small sample of $s=20$ elements. Global
maximum and minimum elements over a set of processes are computed by
an \mpiallreduce operation with the \MPIMAX operator. The global pivot
is interpolated as the average of global maximum and minimum
element. As inputs we have used either random permutations of
$0,1,\ldots,n-1$ or uniformly generated random numbers in the range
$[0,n-1]$. With these inputs, the chosen pivot selection procedure
leads to (almost) perfect pivots, and all processes process almost the
same number of $n/p$ elements throughout. For the standard parallel
Quicksort variants, standard partition with sentinel elements into two
parts is used, but such that sequences of elements equal to the pivot
are evenly partitioned. For inputs with many equal elements, partition
into three segments might be used to improve the load
balance~\cite{BentleyMcIlroy93}. For the HyperQuicksort variants, the
multiway merge is done using a straightforward binary heap, see,
\eg~\cite{SedgewickWayne11}. For both variants without element
exchange, the final data redistribution is done by an \mpialltoall
followed by an \mpialltoallv operation\footnote{The implementations
  are available from the author.}.

\begin{figure}
  \includegraphics[width=0.48\textwidth]{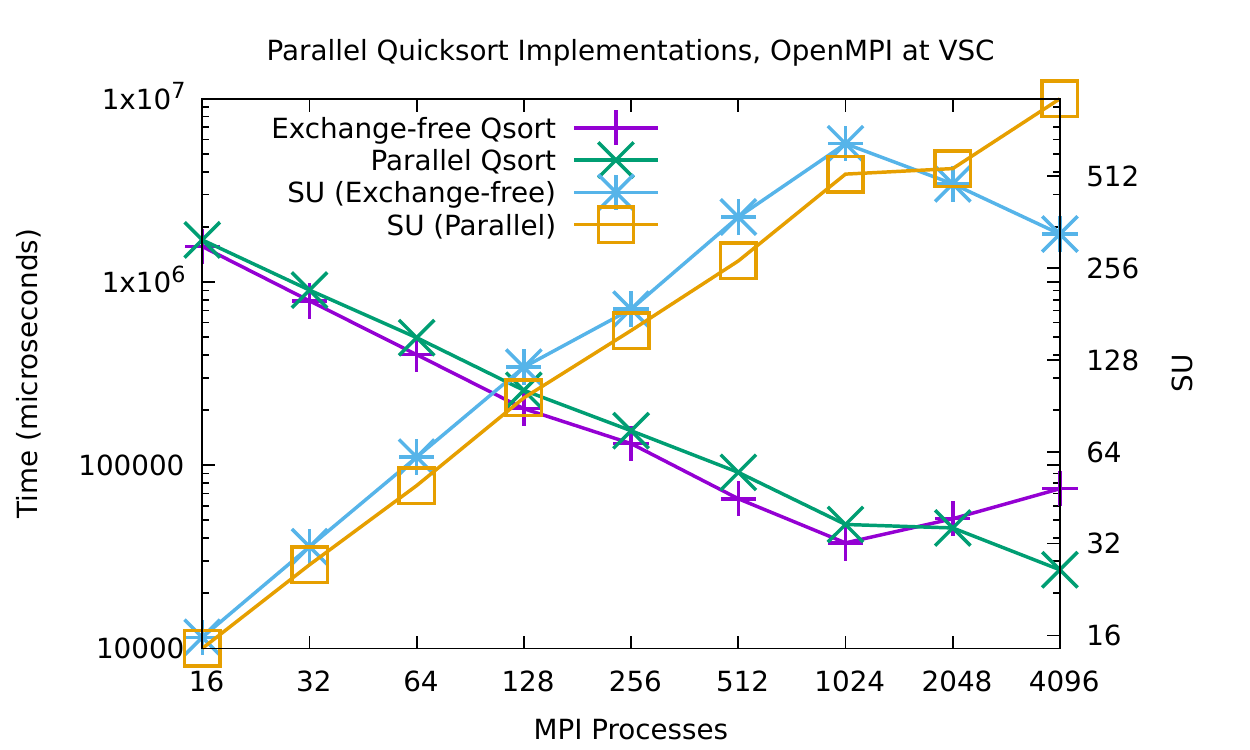}
  \includegraphics[width=0.48\textwidth]{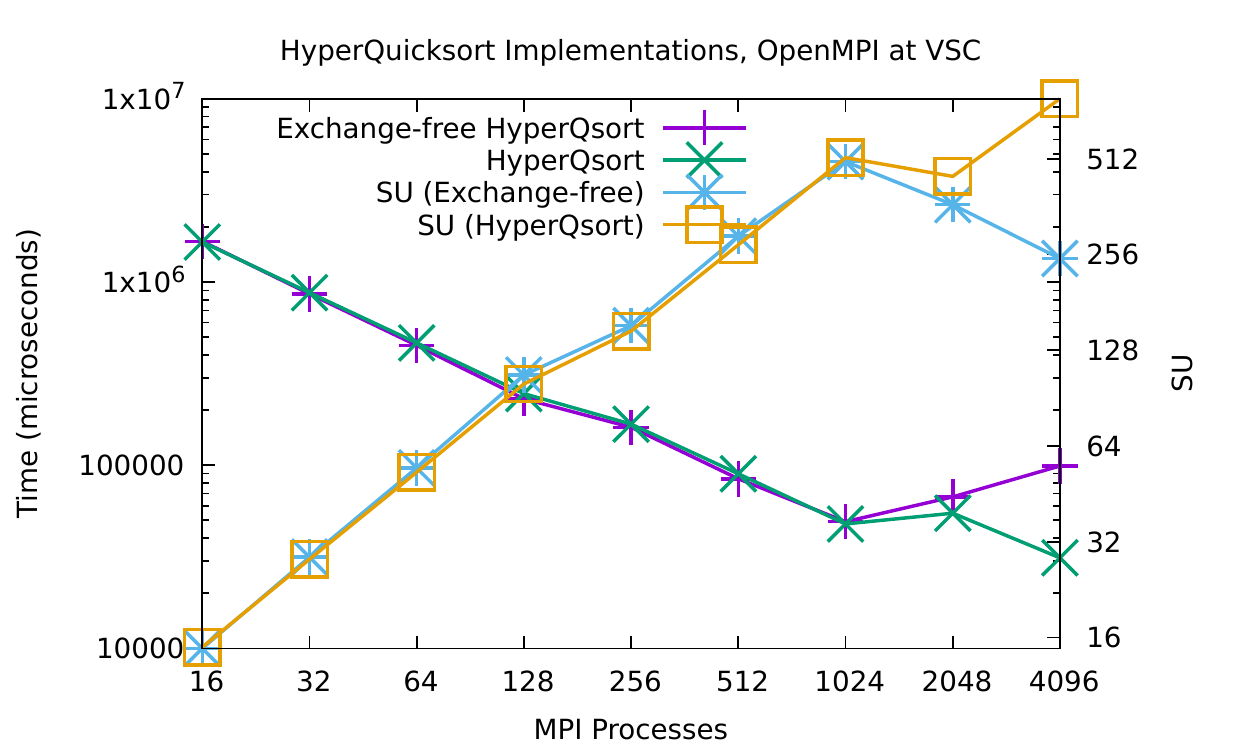}
  \caption{Strong scaling results, $n=10^8$ random doubles in the
    range $[0,10^8-1]$ on a medium-large InfiniBand cluster. Plotted
    running times and speed-up (SU) are the best observed over 43
    measurements. Left plot shows parallel Quicksort versus
    exchange-free parallel Quicksort.  Right plot shows HyperQuicksort
    versus exchange-free HyperQuicksort.}
\label{fig:strongresults}
\end{figure}

The plots in Figure~\ref{fig:strongresults} shows a few strong scaling
results on a medium-sized InfiniBand cluster consisting of 2020
dual-socket nodes with two Intel Xeon E5-2650v2, 2.6 GHz, 8-core Ivy
Bridge-EP processors and Intel QDR-80 dual-link high-speed InfiniBand
fabric\footnote{This is the Vienna Scientific Cluster (VSC). The
  author thanks for access and support.}. The MPI library used is
\openmpilib, and the programs were compiled with \texttt{gcc 6.4} with
optimization level \texttt{-O3}. The $n=10^8$ input elements are
uniformly randomly generated doubles in the range from $[0,10^8-1]$,
and $p$ varies from $2^4$ to $2^{12}$. The $p$ MPI processes are
distributed with 16 processes per compute node. For each input, measurements
were repeated 43 times with 5 non-timed, warm-up measurements, and
the best observed times for the slowest processor are shown in the
plots and used for computing speed-up. As can be seen, the
exchange-free variant of parallel Quicksort gives consistently higher
speed-up by about 20\% than the corresponding standard variant with
element exchanges up to about $1024$ processes. From then on, the
number of elements per process of $n/p<100\,000$ becomes so small that
the \mpiallreduce on the vectors of pivots and the linear latency of
the \mpialltoallv operation become more expensive than the explicit
element exchanges. The exchange-free variant of HyperQuicksort is
slightly faster than HyperQuicksort, but by a much smaller
margin. Interestingly, with (almost) perfect pivot selection as in
these experiments, standard parallel Quicksort seems preferable
to Wagar's HyperQuicksort.

\begin{figure}
  \includegraphics[width=0.48\textwidth]{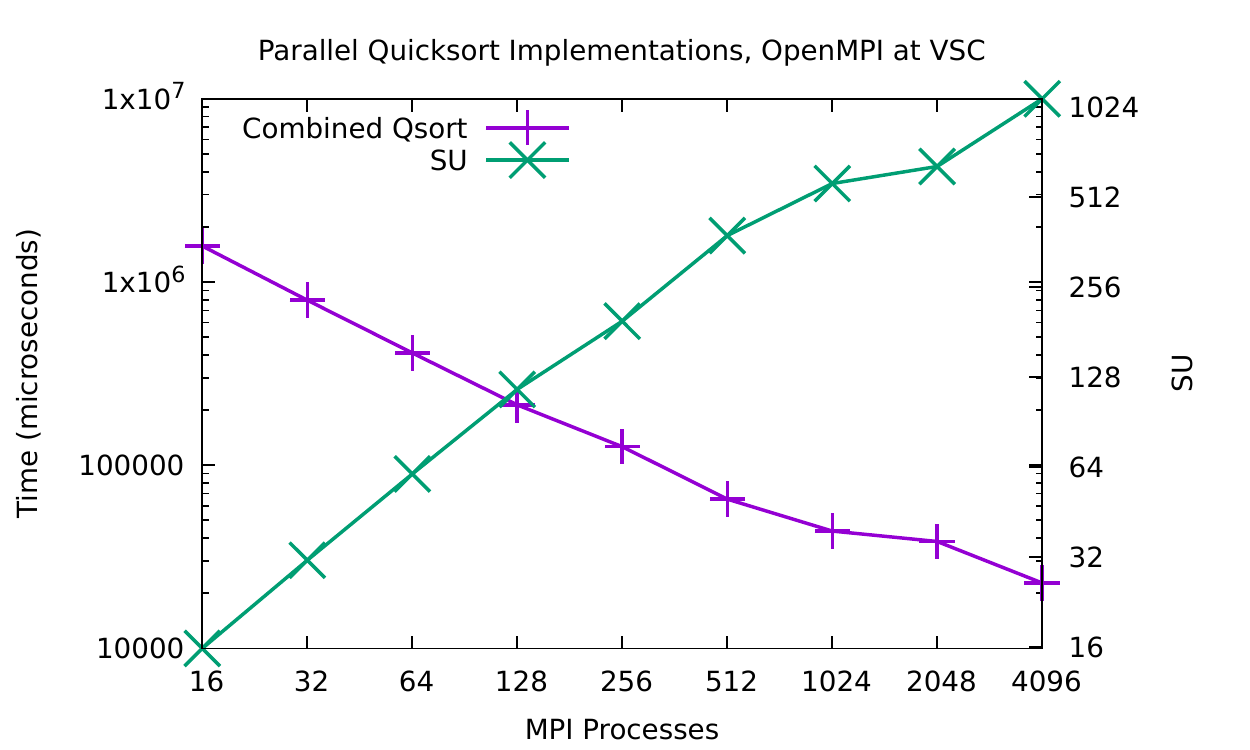}
  \includegraphics[width=0.48\textwidth]{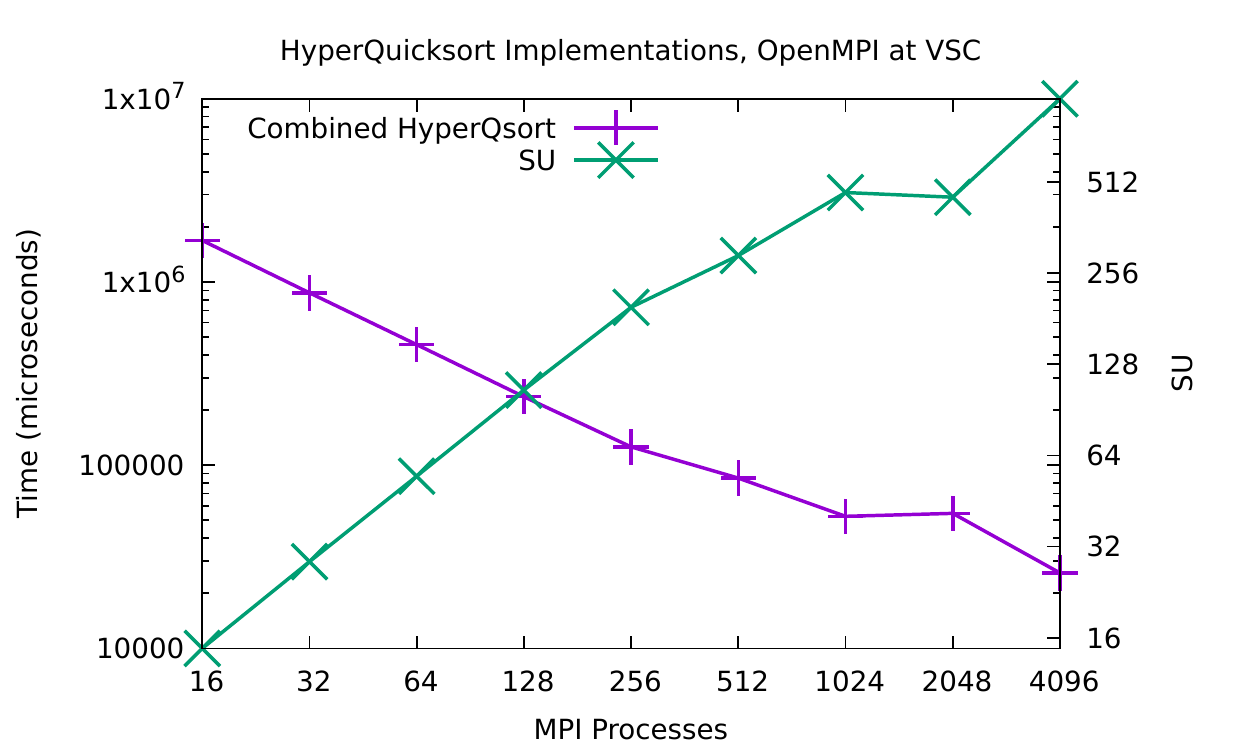}
  \caption{Strong scaling results, $n=10^8$ random doubles in the
    range $[0,10^8-1]$ on a medium-large InfiniBand cluster for the
    combined Quicksort implementation. Plotted running times and
    speed-up (SU) are the best observed over 43 measurements. Left
    plot shows combined, parallel Quicksort. Right plot shows
    combined HyperQuicksort.}
\label{fig:strongcombinedresults}
\end{figure}

To counter the degradation in the exchange-free variants for small
$n/p$, the explicit exchange and exchange-free variants can be
combined. Throughout the recursion in parallel Quicksort and
HyperQuicksort, the number of elements per process $n'$ stays the same
(under the assumption that optimal pivots are chosen) whereas the
number of processes is halved in each recursive call. Thus, the
recursion is stopped when $n'>c p/\log n$ for some chosen
implementation and system dependent constant $c$, and the
corresponding exchange-free variant invoked. By choosing $c$ well,
this will give a smoother transition from parallel Quicksort for small
per process input sizes to exchange-free Quicksort as $n/p$
grows. Results from these \emph{combined Quicksort} variants are shown
in Figure~\ref{fig:strongcombinedresults}. With the constant $c$
chosen experimentally as $c=1500$, the combined Quicksort is never
worse than neither standard nor exchange-free variant. Combined
parallel Quicksort reaches a speed-up of more than 1000 on $p=4096$
processes. This speed-up, and the speed-up of 650 with $p=2048$
processes is larger than achieved with either parallel Quicksort and
exchange-free Quicksort alone.

\begin{figure}
  \includegraphics[width=0.48\textwidth]{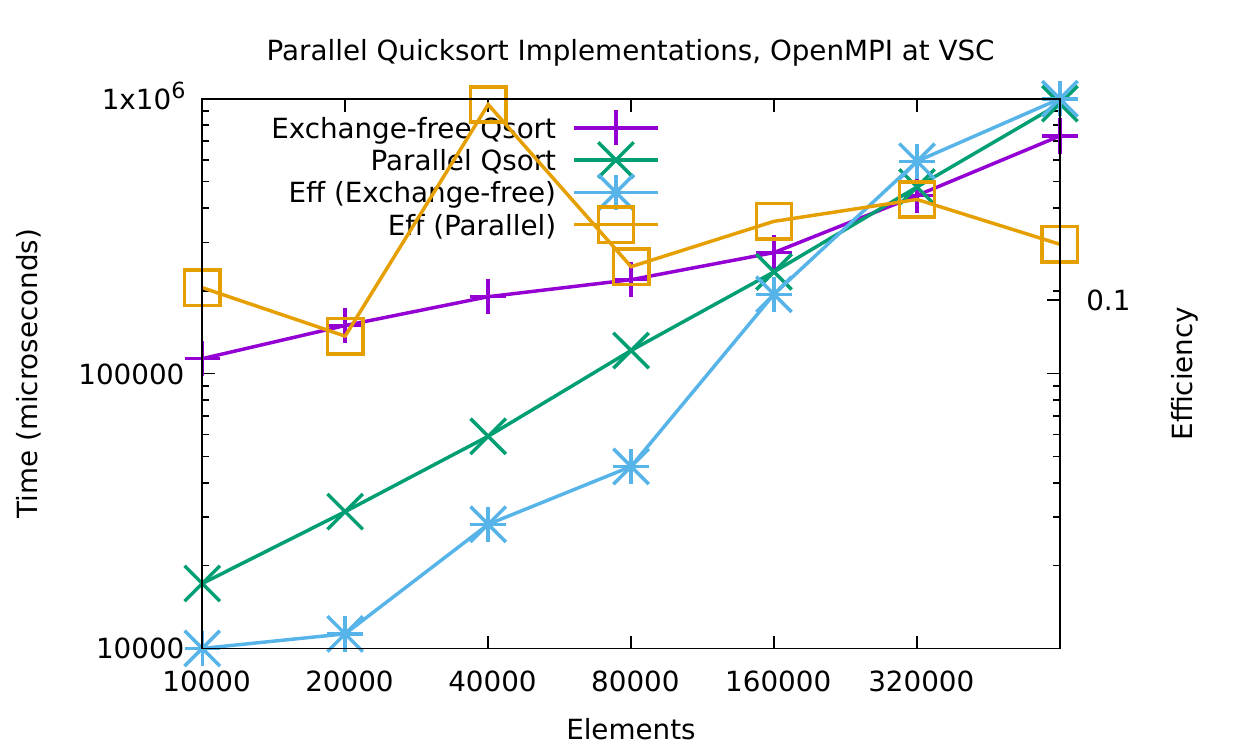}
  \includegraphics[width=0.48\textwidth]{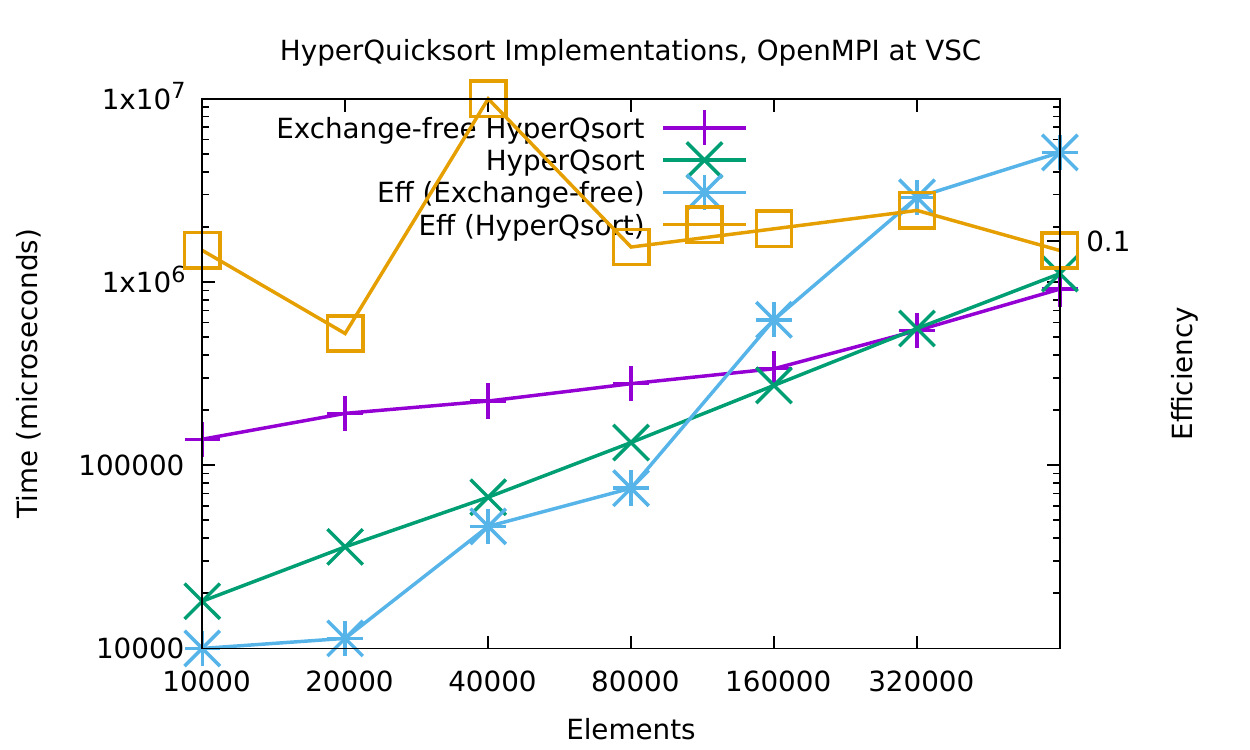}
  \caption{Weak scaling results on a medium-large InfiniBand cluster,
    $p=8192$, and number of elements per process varying from
    $n/p=10\,000$ to $n/p=640\,000$. The input elements are random
    doubles in the range $[0,10^7-1]$. Plotted running times and
    parallel efficiency are the best observed over 43
    measurements. Left plot shows parallel Quicksort versus
    exchange-free parallel Quicksort.  Right plot shows HyperQuicksort
    versus exchange-free HyperQuicksort.}
\label{fig:weakresults}
\end{figure}

The plots in Figure~\ref{fig:weakresults} show results from a weak
scaling experiment where $p=8192$ is kept fixed, and the initial input
size per process varies from $n/p=10\,000$ to $n/p=640\,000$ randomly
generated doubles in the range $[0,10^7-1]$. The experimental setup is
as for the strong scaling experiment. Beyond about $n/p=200\,000$
elements per process, the exchange-free variants perform better than
the standard variants, reaching a parallel efficiency of $15\%$ for
exchange-free Quicksort, and $12\%$ for exchange-free HyperQuicksort.

Repeating the experiments with random permutations and with integer
type elements does not qualitatively change the results.

\section{Concluding remarks}

This note presented two new variants of parallel Quicksort for the
classroom that trade explicit element exchanges throughout the
Quicksort recursion against global selection of multiple pivots and a
single element redistribution. All communication in the new variants
is delegated to MPI collective operations, and the quality of the MPI
library will co-determine the scalability of the implementations.  For
moderately large numbers of elements per process, these variants can
be faster than standard parallel Quicksort variants by a significant
factor, and can be combined with the standard, exchange-based variants
to provide a smoothly scaling parallel Quicksort implementation.

\bibliographystyle{abbrv}
\bibliography{traff,parallel} 

\leaveout{
\appendix

\section{Standard parallel Quicksort algorithms}

For completeness, this appendix gives pseudo-code for the two standard
parallel Quicksort implementation variants, shown as
Algorithm~\ref{alg:standardqsort} (parallel Quicksort) and
Algorithm~\ref{alg:sortfirstqsort}
(HyperQuicksort). Algorithm~\ref{alg:standardqsort} also shows how
standard, parallel Quicksort can transition to exchange-free Quicksort
when the input to be sorted raises above threshold relative to the
number of processes employed. This is the \emph{combined Quicksort}
discussed in the main text. In order that all processes switch
consistently, the maximum number of elements at any process is used as
basis for the decisions. This can either be computed by a collective
\mpiallreduce operation, or an identical estimate for all processes
can be used.

\begin{algorithm}
\caption{Standard parallel Quicksort of elements in $n$-element array $a$ 
for process $i\in\{0,\ldots,p-1\}$ and $p=2^k$.}
\label{alg:standardqsort}
\begin{algorithmic}[1]
\Procedure{ParQsort}{$a,n,i\in\{0,\ldots,p-1\}$}
\If{$p=1$}
\State \Return \Call{local-Qsort}{$a,n$}\Comment{Sort locally}
\ElsIf{$\max_{i\in\{0,\ldots,p-1\}}(n)>c p/(\log_2p+\log_2\max_{i\in\{0,\ldots,p-1\}}(n))$}
\State \Return \Call{ExchangeFreeQsort}{$a,n$}\Comment{Sort locally}
\Else
\State $x\gets\mbox{\Call{local-Choose}{$a,n$}}$\Comment{Local pivot selection}
\State $x'\gets\mbox{\Call{global-Choose}{$x$}}$\Comment{Global consensus}
\State $n_0\gets\mbox{\Call{Partition}{$a,n,x'$}}$
\State \Comment{$a[0,n_0-1]\leq x', a[n_0,n-1]\geq x'$}
\If{$i<p/2$}
\State \Call{Exchange}{$a+n_0,n-n_0,i+p/2,a',n'$}
\State \Call{copy}{$a,n_0,a'+n'$}
\State \Comment{Recursively sort new array $a'$ in parallel}
\State \Return \Call{ParQsort}{$a',n'+n_0,i\in\{0,\ldots,p/2-1\}$}
\Else
\State \Call{Exchange}{$a,n_0,i-p/2,a',n'$}
\State \Call{copy}{$a+n_0,n-n_0,a'+n'$}
\State \Comment{Recursively sort new array $a'$ in parallel}
\State \Return \Call{ParQsort}{$a',n'+n-n_0,i-p/2\in\{0,\ldots,p/2-1\}$}
\EndIf
\EndIf
\EndProcedure
\end{algorithmic}
\end{algorithm}

\begin{algorithm}
\caption{HyperQuicksort of elements in $n$-element array $a$ 
for process $i\in\{0,\ldots,p-1\}$ for $p=2^k$.}
\label{alg:sortfirstqsort}
\begin{algorithmic}[1]
\Procedure{HyperQsort}{$a,n,i\in\{0,\ldots,p-1\}$}
\State \Call{local-Qsort}{$a,n$}\Comment{Sort locally}
\State \Call{MergePartition}{($a,n,i\in\{0,\ldots,p-1\}$)}
\EndProcedure

\Procedure{MergePartition}{$a,n,i\in\{0,\ldots,p-1\}$}
\If{$p=1$} 
\State \Return $a$
\EndIf
\State $x\gets \mbox{\Call{local-Choose}{$a,n$}}$\Comment{Local pivot selection $x=a[n/2]$}
\State $x'\gets\mbox{\Call{global-Choose}{$x$}}$\Comment{Global consensus}
\State $n_0\gets\mbox{\Call{Split}{$a,n,x'$}}$\Comment{$a[0,n_0-1]\leq x', a[n_0,n-1]\geq x'$}
\If{$i<p/2$}
\State \Call{Exchange}{$a+n_0,n-n_0,i+p/2,a',n'$}
\State \Call{Merge}{$a',n',a,n_0,b$}
\State \Return \Call{MergePartition}{$b,n'+n_0,i\in\{0,\ldots,p/2-1\}$}
%\Comment{Recursively sort new array $a'$}
\Else
\State \Call{Exchange}{$a,n_0,i-p/2,a',n'$}
\State \Call{Merge}{$a',n',a+n_0,n-n_0,b$}
\State \Return \Call{MergePartition}{$b,n'+n-n_0,i-p/2\in\{0,\ldots,p/2-1\}$}
%\Comment{Recursively sort new array $a'$}
\EndIf
\EndProcedure
\end{algorithmic}
\end{algorithm}

} % end of appendix

\end{document}